\documentclass[prb,aps,twocolumn,superscriptaddress]{revtex4-1}
\usepackage{graphicx,color}
\usepackage{amsthm}
\usepackage{amsfonts}
\usepackage{algorithmic}
\usepackage{enumerate}
\usepackage{latexsym}
\usepackage{amsmath}
\usepackage{amssymb}
\usepackage{graphicx}
\usepackage[colorlinks=true,citecolor=blue,linkcolor=blue]{hyperref}

\def\avg#1{\left\langle#1\right\rangle}

\emergencystretch=\maxdimen
\begin{document}

\title{Strong ferromagnetic fluctuations in a doped checkerboard lattice}
\author{Yue Pan}
\affiliation{Department of Physics, Beijing Normal University, Beijing 100875, China}
\author{Runyu Ma}
\affiliation{Department of Physics, Beijing Normal University, Beijing 100875, China}
\affiliation{Beijing Computational Science Research Center, Beijing 100193, China}
\author{Tianxing Ma}
\email{txma@bnu.edu.cn}
\affiliation{Department of Physics, Beijing Normal University, Beijing 100875, China}
\affiliation{Key Laboratory of Multiscale Spin Physics(Ministry of Education), Beijing Normal University, Beijing 100875, China\\}

\begin{abstract}
Using the determinant quantum Monte Carlo method, we study the magnetic susceptibility
in the parameter space of the on-site interaction $U$,
temperature $T$, electron filling $\avg{n}$, and the frustration control
parameter $t^{\prime}$ within the Hubbard model on a two-dimensional checkerboard lattice.
It is shown that the system exhibits stable and strong ferromagnetic fluctuations
about the electron filling $\avg{n}\ge1.2$ for different $t^{\prime}$, and the ferromagnetic susceptibility is strongly enhanced by the increasing interaction and decreasing tempeture.
We also discuss the sign problem to clarify which parameter region is accessible and reliable.
Our findings not only demonstrate important implications
for modulating magnetism in the checkerboard lattice, but will also provide a
theoretical platform for a flat-band model that demonstrates a variety of physical properties.
\end{abstract}
\maketitle

\section{Introduction}
Because of the discovery of doped graphene-based materials\cite{vcervenka2009room} and metal-doped transition-metal dichalcogenides (TMDs)\cite{Reyntjens_2021}, which
have the capacity to control magnetic order through charge transfer, the possibility of manufacturing new magnetic devices has attracted much attention.
Researchers are interested in semiconducting two-dimensional (2D) materials doped with impurity atoms, which are considered as a promising platform for high-performance spintronic devices and sensors.
Magnetism plays an important role in the
properties of these materials\cite{PhysRevLett.96.197001}, and in the past few years, people have had great interest in the properties of quantum magnets with geometric frustration\cite{ramirez2003geometric}, such as organic charge
transfer salts of triangular lattices\cite{lunkenheimer2012multiferroicity}, compounds of kagome lattices\cite{PhysRevLett.121.096401,PhysRevX.11.031050,PhysRevLett.127.177001,PhysRevLett.127.217601} and so on, which are expected to describe novel quantum states and interesting
magnetic phases in this field, including quantum spin liquid\cite{PhysRevResearch.1.032011},
and spin ice\cite{balents2010spin,PhysRevB.67.235102,PhysRevLett.89.226402}.
Additionally, pyrochlore oxides like LiV$_{2}$O$_{4}$\cite{PhysRevLett.113.236402,PhysRevB.104.245104}
and Sn$_{2}$X$_{2}$O$_{7}$(X=Nb,Ta)\cite{PhysRevMaterials.1.021601,PhysRevLett.120.196401}
represent another geometric frustration structure, the checkerboard lattice\cite{PhysRevLett.114.130601},
which attracts intensive studies due to its rich phase diagram induced by electronic correlation, including the quantum Hall effect\cite{PhysRevB.90.081102}, superconductivity\cite{santos2010two}, Mott physics\cite{PhysRevB.94.155119,PhysRevB.64.085102}, and other phenomena.
One promising candidate is Sn$_{2}$X$_{2}$O$_{7}$(X=Nb,Ta),
which shows possible ferromagnetism induced by the quasi flat band\cite{PhysRevLett.120.196401}.

Most of previous theoretical investigations primarily focused on the
checkerboard lattice at half-filling. Based on the Heisenberg model with spin
exchange coupling interaction\cite{PhysRevB.95.014420,PhysRevB.99.085112}, rich magnetic characteristics like valence-bond crystal phases\cite{PhysRevB.83.134431}, various magnetization plateaus\cite{PhysRevB.94.140404} are proposed.
The ground-state properties of the geometrically frustrated Hubbard model on the anisotropic
checkerboard lattice at half filling has been studied by using the path-integral renormalization group method\cite{YOSHIOKA2007873,doi:10.1143/JPSJ.77.104702,PhysRevB.78.165113}. It was found that the increase of the Coulomb interaction may induce the first order
metal-insulator transition to the antiferromagnetically ordered phase\cite{YOSHIOKA2007873}, and the plaquette-singlet insulator may emerge besides the
antiferromagnetic insulator and the paramagnetic metal, depending on the frustration-control parameter\cite{PhysRevB.78.165113}.
However, intriguing physical phenomenon always arise as the system is doped.
For simple square lattice, it is well known that it is an Mott insulator with antiferromagnetic $N\acute{e}el$ order for
all values of $U > 0$\cite{Claveau_2014,PhysRevB.50.8039,PhysRevB.51.7038,PhysRevLett.77.4938}.
By using various approaches, including determinant quantum
Monte Carlo (DQMC)\cite{PhysRevB.40.506,PhysRevB.31.4403,PhysRevB.63.125116}, variational
Monte Carlo (VMC)\cite{PhysRevB.38.931,PhysRevB.70.054504}, dynamic cluster approximation
(DCA)\cite{PhysRevB.64.195130}, there are evidences that the doped square lattice exhibit many of the basic physical
properties which characterize the unconventional superconductors\cite{hanaguri2004checkerboard,PhysRevB.91.165109},
for example, antiferromagnetic spin fluctuations\cite{PhysRevLett.124.017003,PhysRevX.11.031007}, pseudogap\cite{PhysRevLett.86.139,PhysRevX.11.021054,PhysRevLett.93.147004},
nematic correlations\cite{PhysRevB.84.220506}, as well as stripes\cite{RevModPhys.84.1383}.
Besides these, doping dependent metal-insulator transition in disordered Hubbard model on a square lattice\cite{PhysRevLett.83.4610,PhysRevB.105.045132}, possible controllability of ferromagnetism in doped honeycomb lttice\cite{doi:10.1063/1.3485059} are also proposed.
These stimulate us to investigate the properties of checkerboard lattice at finite doping,
especially the evolution of magnetic characteristics. Since the checkerboard lattice contains frustration and flat band,
which square lattice does not possess, it would be an interesting topic that how strong ferromagnetic fluctuations emerge in a doped Hubbard model on a checkerboard lattice.

\begin{figure}[tbp]
    \includegraphics[scale=0.35]{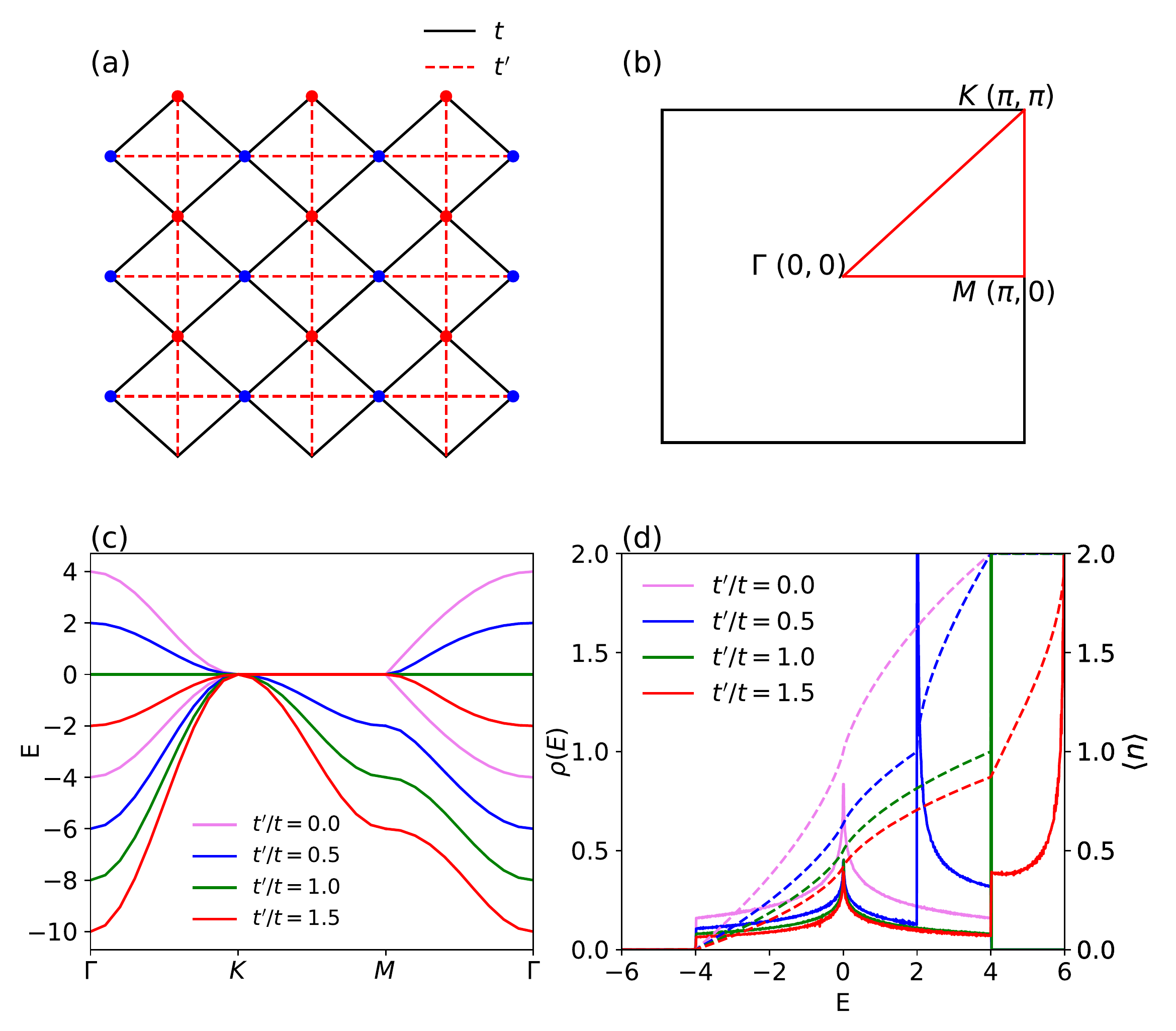}
    \caption{(Color online)(a) Sketch of checkboard lattice. (b) The first
    Brillouin zone and the high symmetry direction (red line). (c) The
    energy band along the high symmetry direction.
    and (d) DOS(solid lines) and filling $n$(dash lines)as functions of energy.}
    \label{Fig1}
    \end{figure}

In the checkerboard lattice, there are two
nonequivalent sites per unit cell of the square lattice,
defining two interpenetrating sublattices, as that illustrated
in Fig. \ref{Fig1}(a), where $t$ represents the nearest hoping and $t'$ indicating the next-nearest hoping.
Fig. \ref{Fig1}(b) shows the high symmetry lines of the first Brillouin zone.
In Fig. \ref{Fig1}(c), one can see that the checkboard lattice exhibits an energetically flat band that is in touch with a quadratically dispersive band along with high symmetry lines of the first Brillouin zone in $k$ space.
The density of states(DOS) are shown in Fig. \ref{Fig1}(d), and there are van Hove singularity in its shape at different fillings depends on the value of $t'$.
Usually, the interplay between electronic correlation and flat band, or van Hove singularity,  may lead to strong ferromagnetic fluctuations.
The magnetic exchange between local spins is largely dominated by the contribution of the flat band, which is ferromagnetic, as revealed by many studies\cite{PhysRevB.103.165105,PhysRevB.85.205104,liu2019flat,PhysRevB.104.155129,PhysRevB.106.155142}.

In this work, we will further provide intensive numerical simulation on the magnetism of doped Hubbard model on a checkerboard lattice by using DQMC method. The checkerboard lattice
that we construct can be realized and modulated through considering next-nearest-neighbor hopping $t^{\prime}$ as frustration-control parameter, to study the effect of the geometric frustration and doping on the magnetic
order at finite temperatures, and interpret the observed magnetic behavior as a result of electronic correlation and the synergetic effect of a special lattice geometry. We also discuss the
sign problem to determine which parameter regions are accessible and unreliable. Interestingly, we found that the system showed obvious
ferromagnetism when the electron filling is $n\ge1.2$, and it is stable and strongest at approximately $n\approx1.5$. The strong ferromagnetic fluctuations we revealed can be understood in the framework of the flat band scenario.

\section{Model and methods}
We consider the Hubbard model defined on a two-dimensional checkerboard lattice.
The checkerboard lattice possesses two sublattices, so the total lattice sites number is
$2\times L^2$. The Hamiltonian that we studied is given by

\begin{equation}
\begin{aligned}
\mathcal{H} &= \mathcal{H}_{1} + \mathcal{H}_{2} + \mathcal{H}_{3} + \mathcal{H}_{4} \\
\mathcal{H}_{1} &= -t\sum_{\langle i, j \rangle\sigma} \left( c^{\dagger}_{iA\sigma} c_{jB\sigma} + H.C. \right)\\
\mathcal{H}_{2} &= -t^{\prime}\sum_{i\sigma} \left( c^{\dagger}_{iA\sigma} c_{i+xA\sigma} + c^{\dagger}_{iA\sigma} c_{i-xA\sigma} + H.C. \right) \\
\mathcal{H}_{3} &= -t^{\prime}\sum_{j\sigma} \left( c^{\dagger}_{iB\sigma} c_{i+yB\sigma} + c^{\dagger}_{iB\sigma} c_{i-yB\sigma} + H.C. \right) \\
\mathcal{H}_{4} &= U \sum_{il} n_{il\uparrow} n_{il\downarrow} + \mu \sum_{il\sigma} n_{il\sigma}
\end{aligned}
\label{eq:ham}
\end{equation}
where $t$ is the hopping amplitude between the nearest-neighbor sites on the lattice. We set $t=1$ as a unit of energy,
where $t^{\prime}$ is the next-nearest-neighbor (NNN) hopping integral
and $c^{\dagger}_{il\sigma}$($c_{il\sigma}$) indicates the creation (annihilating) of an electron at a site $i$.
The sublattice is $l$ and the spin is $\sigma$. $n_{il\sigma}=c^{\dagger}_{il\sigma}c_{il\sigma}$ is the corresponding
particle number operator and $\langle i, j \rangle$ denotes the nearest neighbor. The second term $H_{2}$
represents the next-nearest hopping term of the A sublattice, where only the hoppings in the $x$ direction are considered.
The third term $H_{3}$ represents the next nearest B sublattice, and only $y$ direction hoppings are considered.
The last term contains the Hubbard interaction $U$, which is the strength
of electron repulsion, and the chemical potential $\mu$, which tunes the electron filling.

We can carry out a Fourier transformation $c_{kl\sigma} = \sum_{j} e^{ijR} c_{jl\sigma}$
to extract the band structure in the noninteracting limit ($U \rightarrow 0$).
After Fourier transformation, the noninteracting
Hamiltonian can be block diagonalized to $\mathcal{H} = \sum_{k} \mathcal{H}(k)$, and $\mathcal{H}(k)$
can be written as

\begin{equation}
    H=\sum_{\mathbf{k}\sigma }(%
    \begin{array}{cc}
    c_{\mathbf{k}1\sigma }^{\dag } & c_{\mathbf{k}2\sigma }^{\dag }%
    \end{array}%
    )(%
    \begin{array}{cc}
    \alpha _{\mathbf{k}} & \beta _{\mathbf{k}} \\
    \beta _{\mathbf{k}} & \gamma _{\mathbf{k}}\
    \end{array}%
    )(%
    \begin{array}{c}
    c_{\mathbf{k}1\sigma } \\
    c_{\mathbf{k}2\sigma }%
    \end{array}%
    ).
    \end{equation}

    where
    \begin{equation}
    \begin{aligned}
    \beta_{k} &= -2t\left[\cos \frac{\mathbf{k}}{2}\cdot\left(\vec{a_1}+\vec{a_2}\right)
    + \cos \frac{\mathbf{k}}{2}\cdot \left(\vec{a_1}-\vec{a_2}\right)\right]\\
    \alpha_{k} &= -2t^{\prime} \cos \mathbf{k} \cdot \vec{a_1}\\
    \gamma_{k} &= -2t^{\prime} \cos \mathbf{k} \cdot \vec{a_2}
    \end{aligned}
    \end{equation}

The unit vector $\vec{a_1}=(1,0)$, $\vec{a_2}=(0,1)$.
Diagonalizing this matrix can transform the Hamiltonian into a band basis, and the two eigenvalues
of this matrix are $E^{\pm}(k) =1/2[\alpha_{k}+\beta_{k}] {\pm} 1/2 \sqrt{\alpha_{k}^{2}-2\alpha_{k}\gamma_{k}+4\beta_{k}^{2}+\gamma_{k}^{2}}$.
The band structure is plotted in Fig. \ref{Fig1} (c), where we can see that the flat band starts to develop when
the $t^{\prime}$ term is introduced, and the upper band becomes completely flat when $t^{\prime}/t=1$.

We perform DQMC\cite{article,PhysRevB.28.4059,PhysRevD.24.2278} simulations to extract the finite temperature properties of a checkerboard lattice
at different fillings and different $t^{\prime}/t$ values. The main principle of the DQMC algorithm is as follows; first, the partition function $\mathcal{Z} = Tr e^{-\beta \mathcal{H}}$
is expressed in a discretized imaginary time slice, and this step is called the Trotter decomposition.
Next, we decouple the interaction term by Hubbard-Stratonovich(HS) transformation\cite{PhysRevB.40.506,PhysRevD.24.2278}. After this step,
an auxiliary field couples to electrons, and the interaction term disappears. Then, we can trace out
the electron freedom and the resulting determinant becomes the weight in the sampling process. The observable
can be written as $\langle O \rangle = \frac{Tr\left[e^{-\beta H} O\right]}{\mathcal{Z}}=\sum_{s} \mathbf{P}_{s} \langle O \rangle _{s}$,
where the weight is $\mathbf{P}_{s} = \frac{det(1 + B_{s}(\beta, 0))}{\sum_{s} det(1+B_{s}(\beta, 0))}$, and in practice,
the sampling used is based on a single-flip algorithm, and the accept ratio is $R=\frac{\mathbf{P}_{s^{\prime}}}{\mathbf{P}_{s}}$.
Green functions of  certain auxiliary field configurations can be computed by the formula
$G^{\sigma} = \left[I + \prod_{l} B^{\sigma}_{l} \right]^{-1}$. The $B$ matrix introduced above is $B^{\sigma}_{l} = e^{\sigma \Delta \tau \lambda s_l} e^{-\Delta \tau H_{0}}$,
and $s_l$ is the auxiliary field introduced, $\lambda$ is the corresponding coefficient, and $H_{0}$ is the noninteracting part of the Hamiltonian.
One should note that after HS transformation, the action becomes bilinear, and the correlations can be calculated
by using the Wick theorem, then all observables can be calculated by Green functions. In practice, we start at a random initialized auxiliary
field, then we conduct a warm-up process without calculating observables. Next, we conduct several measurements to accumulate observables
into bins. In our simulations, we use 8000 sweeps to equilibrate the system and
an additional 10000$\sim$200000 sweeps to generate measurements, which were
split into 10 bins to provide the basis of the coarse-grain averages. The valid of this method has been verified in many previous studies, including doped grahpene\cite{doi:10.1063/1.3485059}, Iron-based superconductor\cite{PhysRevLett.110.107002}, as well as highly geometry frustrated system\cite{PhysRevB.80.014428}.

\begin{figure}[tbp]
\includegraphics[scale=0.35]{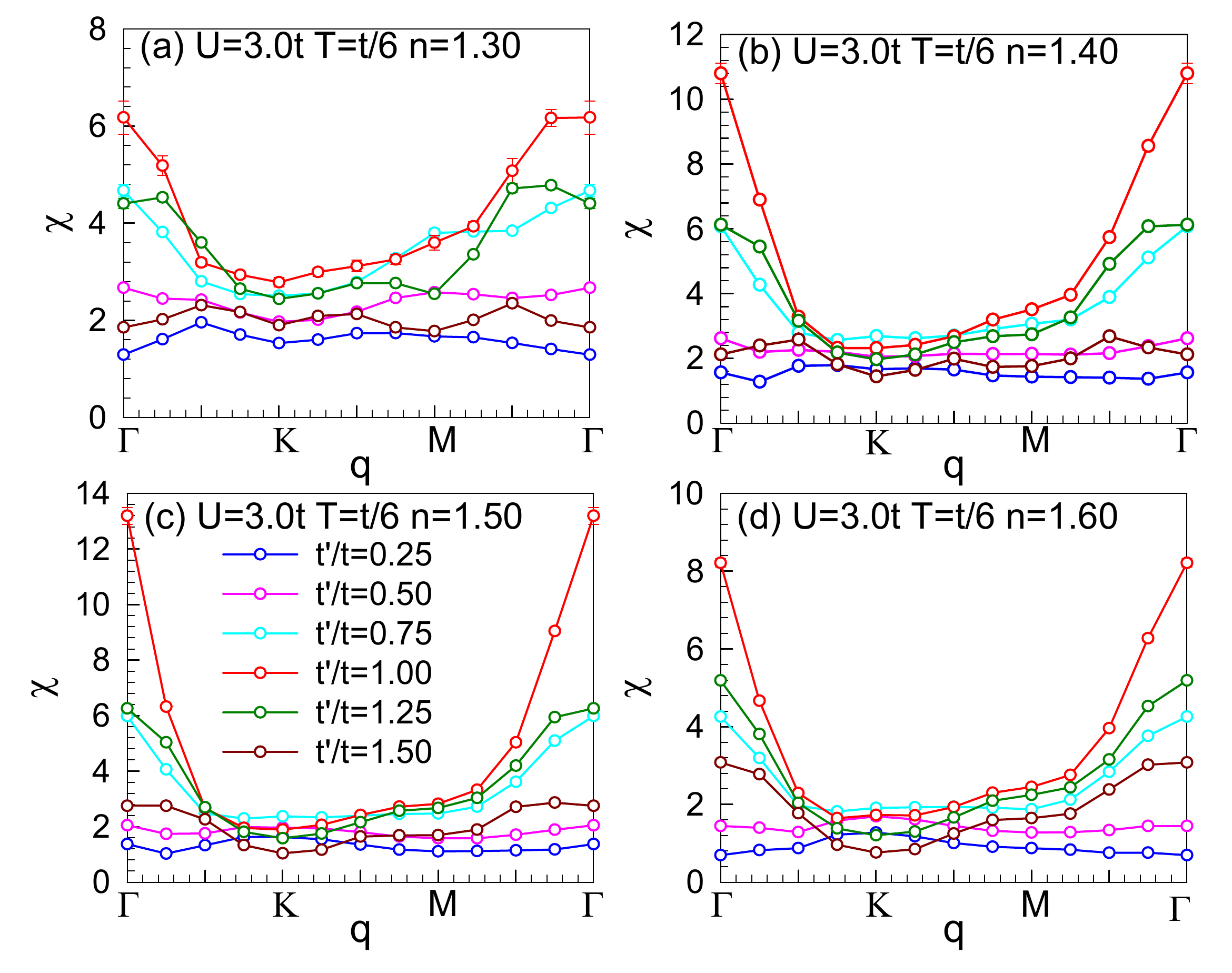}
\caption{(Color online) Magnetic susceptibility vs. momentum $q$ at different $t^{\prime}/t$. Here $U=3.0t$, $T=t/6$ for (a)$n=1.30$, (b)$n=1.40$, (c)$n=1.50$, (d)$n=1.60$.}
\label{Fig2}
\end{figure}

To study ferromagnetic fluctuations, we define the spin susceptibility in the z direction at zero frequency,
\begin{equation}
\chi(q) = \frac{1}{N} \sum_{i, j} \sum_{l, m} \int^{\beta}_{0} d \tau e^{-i q (\mathbf{R}_{i} - \mathbf{R}_{j})}
\left\langle S^{z}_{i, l}(\tau) S^{z}_{j,m}(0) \right\rangle
\end{equation}
where $S^{z}_{i, l}(\tau)=e^{H \tau }S^{z}_{j,m}(0)e^{-H\tau}$ with
$S^{z}_{i, l}=c^{\dagger}_{il\uparrow} c_{il\uparrow}-c^{\dagger}_{il\downarrow}c_{il\downarrow}$.
To have a deep understanding of the effect of $t^{\prime}$ on the magnetic order, we look at the
spin susceptibility as a function of temperature $T$, interaction $U$ and electronic filling $n$ on a $L=8$ lattice for several typical $t^{\prime}/t = 0.25 \sim 1.50$.

\section{Results and discussion}
In Fig. \ref{Fig2}, we show the magnetic susceptibility $\chi(q)$ along with high symmetry lines of the first Brillouin zone including $\Gamma=(0,0)$, $K=(\pi,\pi)$, $M=(\pi,0)$.
We can see that $t^{\prime}/t$ term has a significant effect at $U=3.0t$, $T=t/6$, with different filling (a)$n =1.30$, (b)$n =1.40$, (c)$n =1.50$, and (d)$n =1.60$.
$\chi(\Gamma)$ gets enhanced greatly as $t^{\prime}/t$ increases, while
$\chi(K)$ increases only slightly. The change of $\chi(\Gamma)$ is considered as significant ferromagnetic
fluctuation with increasing $t^{\prime}/t$.
And we may also notice that $\chi(\Gamma)$ have a peak at $t^{\prime}=1$, where the upper band is a flat band.
This reflects the importance of the flat band and electron structure in the shaping of ferromagnetic behavior.

\begin{figure}[tbp]
\includegraphics[scale=0.35]{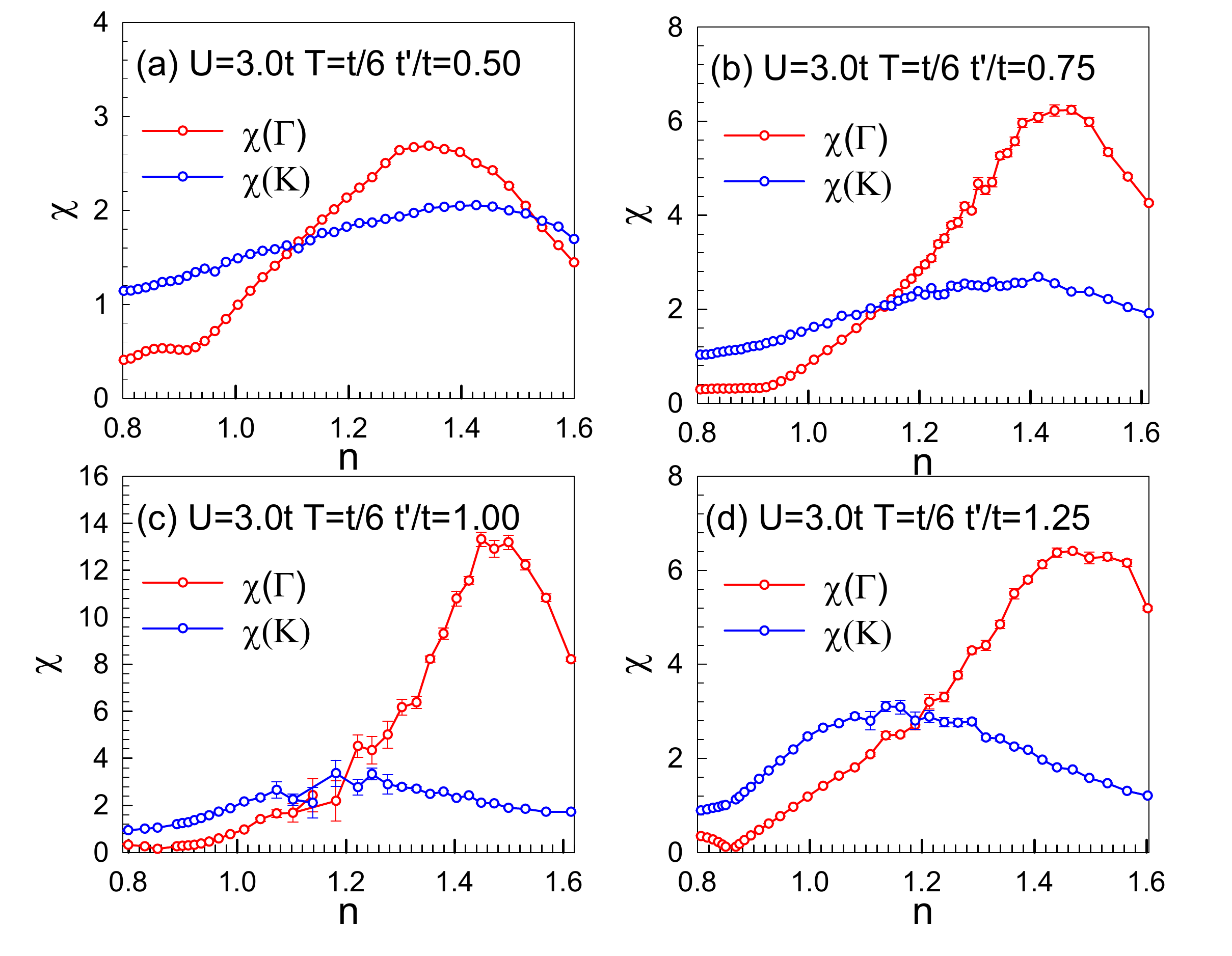}
\caption{(Color online) Magnetic susceptibility $\chi(\Gamma)$ (red) and $\chi(K)$(blue) vs. electron filling $n$ at $U=3.0t$, $T=t/6$ with (a)$t^{\prime}/t=0.50$, (b)$t^{\prime}/t=0.75$, (c)$t^{\prime}/t=1.00$, and (d)$t^{\prime}/t=1.25$.}
\label{Fig3}
\end{figure}
To understand the filling dependence of magnetic correlations in checkerboard lattice,
we show in Fig. \ref{Fig3} that the magnetic susceptibilities for ferromagnetic
fluctuation $\chi(\Gamma)$ and antiferromagnetic fluctuation $\chi(K)$
at $T= t/6$ and $U=3.0t$ with different values of $t^{\prime}/t$, (a)$t^{\prime}/t =0.50$, (b)$t^{\prime}/t =0.75$, (c)$t^{\prime}/t =1.00$, and (d)$t^{\prime}/t =1.25$. 
Here we can see
$\chi(\Gamma)$ increases faster than $\chi(K)$ at filling in the range 1.20 to 1.60, $\chi(\Gamma)$ becomes higher than $\chi(K)$,
indicating that ferromagnetic fluctuation is dominant.
And the maximum of magnetic susceptibilities at approximately $n\approx1.5$, which is stable and the strongest ferromagnetic fluctuation.

\begin{figure}[tbp]
\includegraphics[scale=0.35]{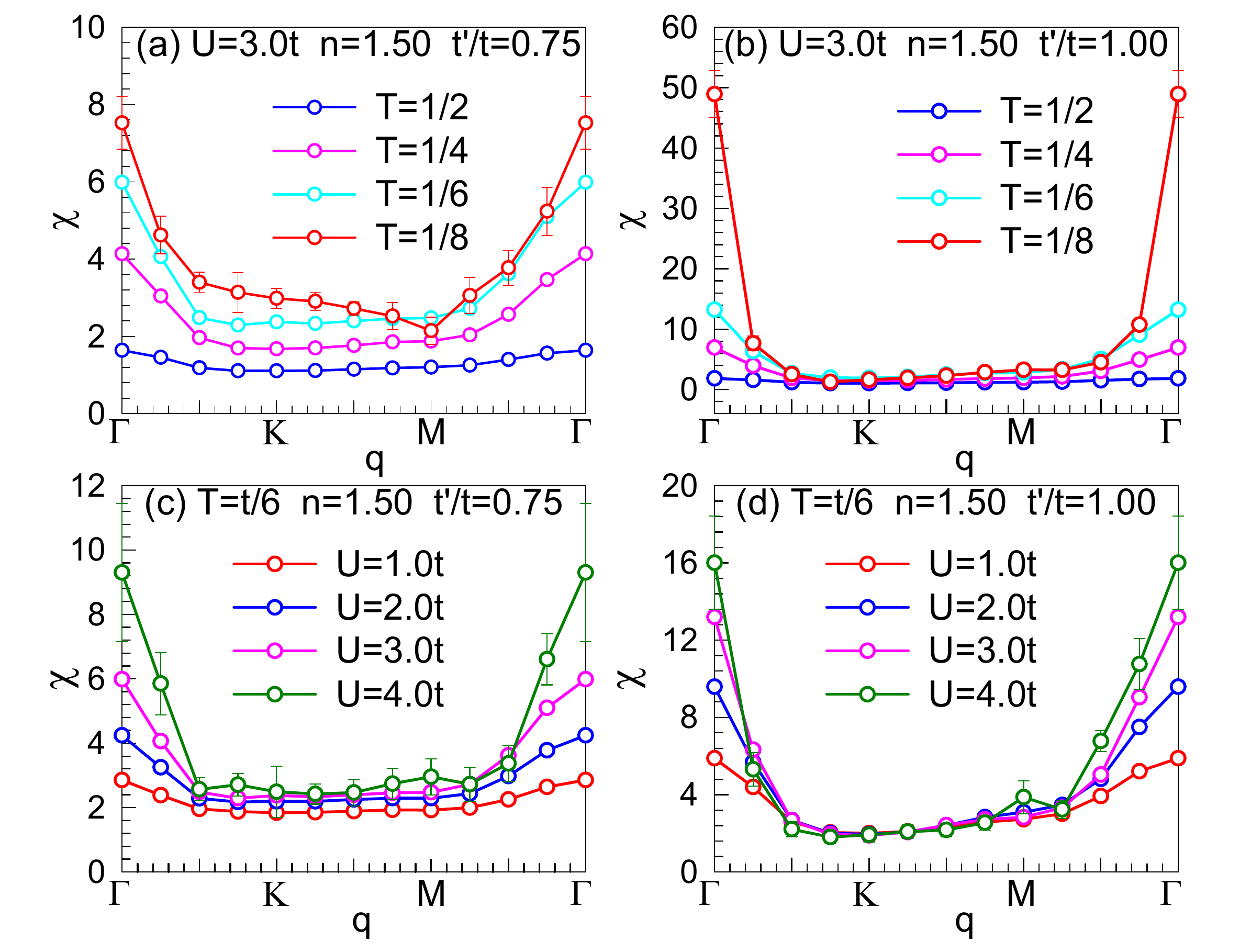}
\caption{(Color online) The magnetic susceptibility vs. the momentum $q$ at different values of $T$ with $U=3.0t$, $n=1.50$, (a)$t^{\prime}/t=0.75$, and (b)$t^{\prime}/t=1.00$.
The magnetic susceptibility vs. the momentum $q$ at different values of $U$ with $T=t/6$, $n=1.50$, (c)$t^{\prime}/t=0.75$, and (d)$t^{\prime}/t=1.00$.}
\label{Fig4}
\end{figure}

Then, in the section of this paper, we investigate the properties of ferromagnetic correlation at a fixed electron filling $n=1.50$.
In Fig. \ref{Fig4}(a) and Fig. \ref{Fig4}(b),
we show the magnetic susceptibility at the filling $n=1.50$, $U=3.0t$, $t^{\prime}/t =0.75$, and $t^{\prime}/t =1.00$.
For temperature $T$ ranging from 1/2 to 1/8, the magnetic susceptibility increases as $T$ decreases.
In Fig. \ref{Fig4}(c) and Fig. \ref{Fig4}(d), we show the magnetic susceptibility at the filling $n=1.50$, $t^{\prime}/t=0.75$, and $t^{\prime}/t=1.00$.
For $U$ ranging from 1.0 to 4.0 and temperature $T= t/6$, the magnetic susceptibility increases as $U$ increases.
This reflects the enhancement of ferromagnetic correlation by interaction $U$.

\begin{figure}[tbp]
\includegraphics[scale=0.4]{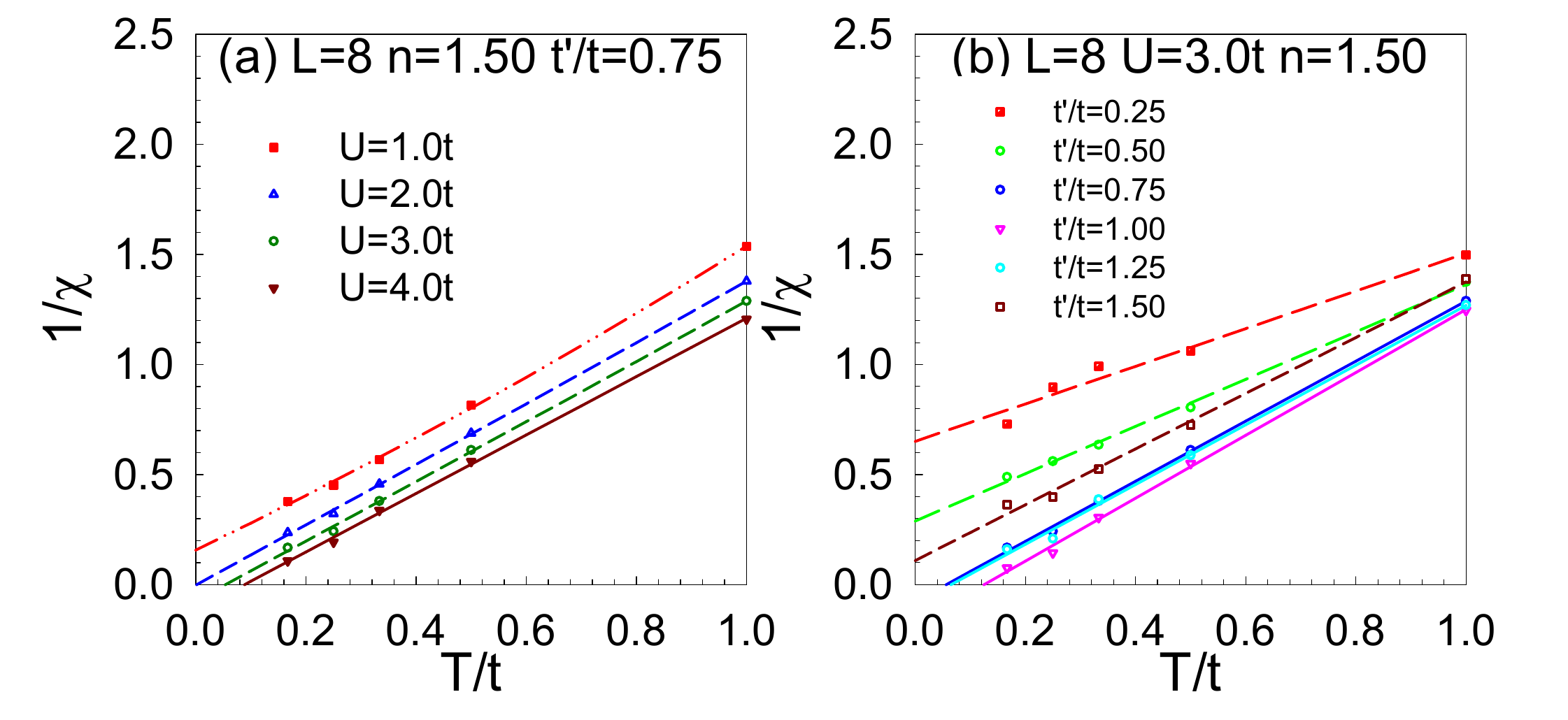}
\caption{(Color online) (a) The temperature dependent $1/\chi$ at $n=1.50$ and $t^{\prime}/t=0.75$ with different $U$.
(b) The temperature dependent $1/\chi$ at $U=3.0t$ \cite{}and $n=1.50$ with different $t^{\prime}/t$.}
\label{Fig5}
\end{figure}

Then, we present the temperature dependence of the magnetic susceptibility at $n=1.50$ with different $U$ and $t^{\prime}/t$
in Fig. \ref{Fig5}(a) and Fig. \ref{Fig5}(b). The figure exhibits a linear correlation between
$1/\chi$ and temperature $T$,
which corresponds to the Cuire-Weiss behavior $1/\chi = (T- T_{c}) / A$.
Therefore, we extrapolate $1/\chi$ to zero temperature by using linear fitting. If the
system possess a finite $T_{c}$, its intercept should be negative.
Fig. \ref{Fig5}(a) shows that the on-site interaction $U$ enhances the ferromagnetic fluctuations and
the negative intercept appears at approximately $U=2.0t$.
Fig. \ref{Fig5}(b) shows that the fully-frustrated case $t^{\prime}/t=1.0$ has the lowest intercept,
which means it has the highest $T_{c}$.

\begin{figure}[tbp]
\includegraphics[scale=0.36]{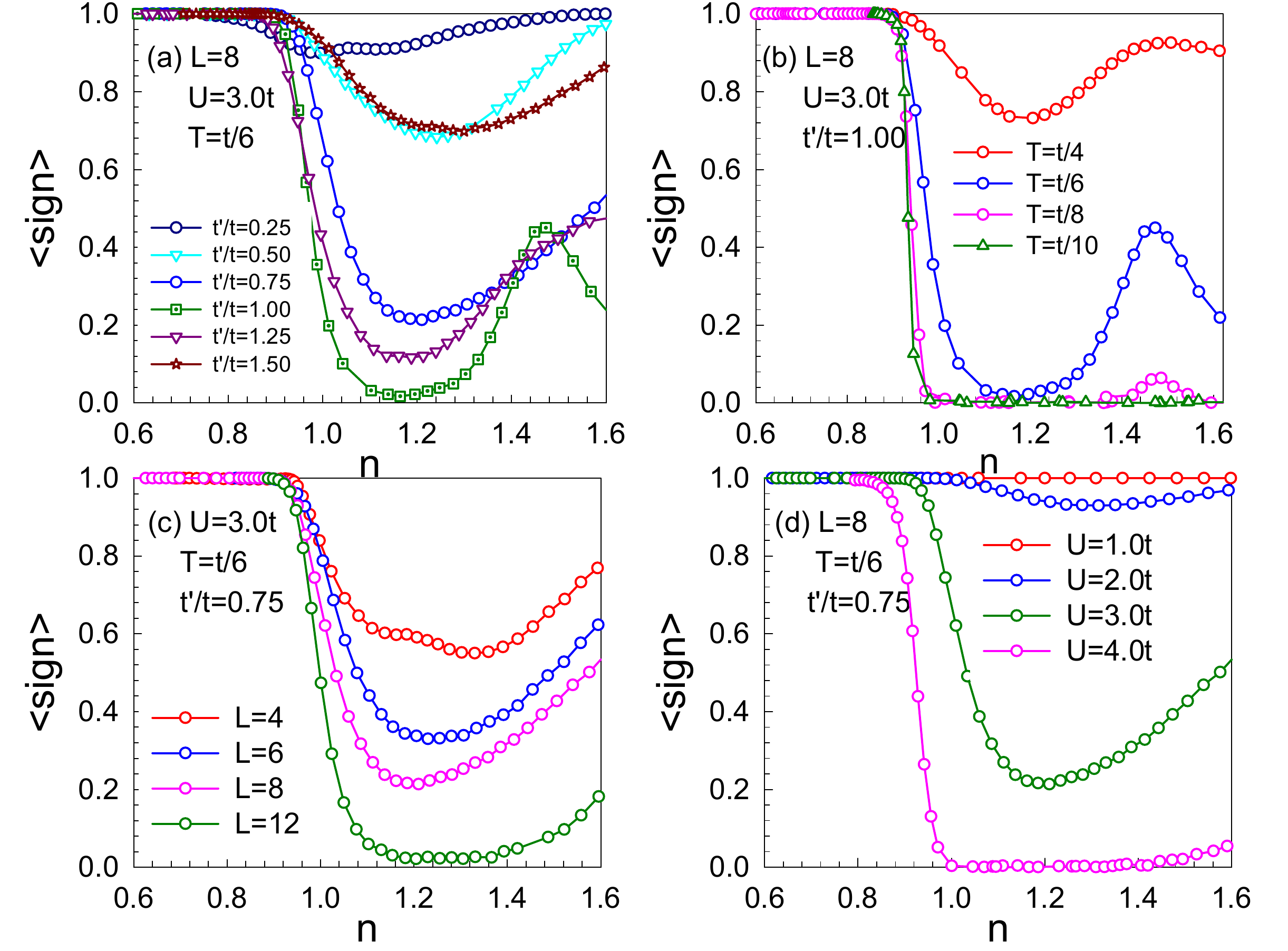}
\caption{(Color online) The average sign as a function of electron
filling $n$ for (a) different $t^{\prime}/t$, (b) different temperature $T$,
(c) different lattice sizes $L$, and (d) different interactions $U$.}
\label{Fig6}
\end{figure}

According to previous studies, the sign problem is a major obstacle to reaching low temperatures and strong-
coupling regions in the QMC simulations. In Fig. \ref{Fig6}, the
average sign evolves with electron filling $n$ while other
parameters are fixed, for doped checkerboard lattice.
In our simulations, especially in the following simulation
results where the sign problem is much worse, we
have increased measurement from 10000$\sim$200000
times to compensate the fluctuations, which is large
enough to ensure the reliability and accuracy of the data\cite{HUANG2019310}.
The four subfigures in Fig. \ref{Fig6} plot the average sign
with (a) different values of $t^{\prime}/t$, (b) different temperatures $T$, (c) different interaction $U$, and (d) different lattice sizes $L$.

In Fig. \ref{Fig6} (a), the sign problem becomes worse along with hopping
strength $t^{\prime}$, which increases at first and then decreases as the
value of $t^{\prime}/t$ continuously increases. Particularly, when $t^{\prime}/t=1.00$, it is shown that the average sign falls to minimum value.
We note another universal feature for all values of $t^{\prime}$, which is that $\langle sign \rangle$ presents
a minimum at electron filling $n$ around $1.20$. As shown in the analysis of magnetism in Fig. \ref{Fig3},
we find that the value where ferromagnetic fluctuations becomes dominant is also $n \approx 1.20$.
It has been reported that $\langle sign \rangle$ can be related to the quantum phase transition\cite{doi:10.1126/science.abg9299},
So, this minimum in the $\langle sign \rangle$ may come out of the ferromagnetism.

In Fig. \ref{Fig6}(b), the average sign decays exponentially with decreasing temperature.
The average sign is almost zero when $n>1.00$ at the temperatures $T=t/8$ and $T=t/10$, making the DQMC simulations nearly impossible.
Moreover, by comparing various values of lattice size $L$ for the $t^{\prime}/t=0.75$ ,
we know that as the lattice size increases, the value of the average sign decreases, as shown in Fig. \ref{Fig6} (c).
In Fig. \ref{Fig6}(d), for values of $U$ in the range $1.0$ to $4.0$ with the temperature fixed at $T= t/6$,
the sign problem becomes worse as $U$ becomes larger.

\section{conclusions}
In this work, we study the finite temperature properties of the
checkerboard lattice Hubbard model by means of the
DQMC method. Our results present exact numerical results
on the magnetic correlation. By controlling the geometric frustration via
a systematic change in the transfer integral $t^{\prime}$ along the diagonal
bonds, we show the wide parameter region from $t^{\prime}/t= 0.25$ to $t^{\prime}/t =1.50$,
including the fully frustrated checkerboard lattice $t^{\prime}/t =1.00$,
where there is a stable ferromagnetic state at electron filling $n=1.5$ for a certain temperature, and this effect is obviously strengthened as
the interaction $U$ increases. We also discuss the sign problem to clarify which parameter regions are accessible and reliable.
Our findings not only have important implications for exploiting emergent flat band
physics in special lattice geometries, but also may shed light on the competition between magnetic mode of highly frustrated systems.
Therefore, we provide further guidance for experiments in this parameter space.

\noindent
\underline{\it Acknowledgement} ---
This work was supported by NSFC (No. 11974049).
The numerical simulations in this work were performed at HSCC of
Beijing Normal University.
\bibliography{reference}

\end{document}